\documentclass[twocolumn,preprintnumbers,amsmath,amssymb, bibnotes,nofootinbib,superscriptaddress]{revtex4}
\usepackage{amsthm}
\usepackage{amsmath}
\usepackage{amssymb}
\usepackage{graphicx}
\usepackage{url}
\usepackage{dsfont}
\usepackage{float}
\usepackage{bm}
\usepackage{ifthen}
\usepackage[usenames,dvipsnames]{color}
\usepackage{mathrsfs}
\usepackage[colorlinks=true,citecolor=blue,urlcolor=black]{hyperref}
\usepackage{float}
\usepackage{afterpage}
\usepackage{subfigure}
\usepackage{adjustbox}
\usepackage{times}

\newcommand{\sket}[1]{{\ensuremath{\lvert#1\rangle}}}
\newcommand{\lket}[1]{{\ensuremath{\left\lvert#1\right\rangle}}}
\newcommand{\ket}[1]{\if@display\lket{#1}\else\sket{#1}\fi}
\newcommand{\sbra}[1]{{\ensuremath{\langle#1\rvert}}}
\newcommand{\lbra}[1]{{\ensuremath{\left\langle#1\right\rvert}}}
\newcommand{\bra}[1]{\if@display\lbra{#1}\else\sbra{#1}\fi}
\newcommand{\sbraket}[2]{{\ensuremath{\langle#1\rvert#2\rangle}}}
\newcommand{\lbraket}[2]{{\ensuremath{\left\langle#1\!\left\rvert\vphantom{#1}#2\right.\!\right\rangle}}}
\newcommand{\braket}[2]{\if@display\lbraket{#1}{#2}\else\sbraket{#1}{#2}\fi}

\newcommand{\sketbra}[2]{{\ensuremath{\lvert #1\rangle\!\langle #2\rvert}}}
\newcommand{\lketbra}[2]{{\ensuremath{\left\lvert #1\right\rangle\!\!\left\langle #2\right\rvert}}}
\newcommand{\ketbra}[2]{\if@display\lketbra{#1}{#2}\else\sketbra{#1}{#2}\fi}

\newcommand{\proj}[1]{\ketbra{#1}{#1}}

\newcommand{\eps}{\epsilon}

\newcommand{\sT}{\mathsf{T}}
\newcommand{\sn}{\mathsf{n}}
\newcommand{\sW}{\mathsf{W}}
\newcommand{\rs}{{\rm{1}}}
\newcommand{\rd}{{\rm{2}}}
\newcommand{\rdd}{{\rm{3}}}

\newcommand{\rvX}{\textbf{X}}

\theoremstyle{plain}

\theoremstyle{definition}

\newcommand*{\cH}{\mathcal{H}}

\newcommand*{\cT}{\mathcal{T}}

\newcommand*{\cW}{\mathcal{W}}

\newcommand\ddo[1]{\ensuremath{\frac{ \mathrm{d} #1}{2\pi}}}

\usepackage{color}
\definecolor{myred}{rgb}{1,0,0}

\definecolor{myblue}{rgb}{0,0,0.8}

\definecolor{myyellow}{rgb}{0.9,0.8,0}

\definecolor{mygreen}{rgb}{0,0.6,0}

\definecolor{myorange}{rgb}{0.6,0.6,0}

\definecolor{mycerul}{rgb}{0,0.6,1}

\newcommand{\tn}[1]{\textnormal{#1}}
\newcommand{\be}{\begin{equation}}
\newcommand{\ee}{\end{equation}}

\newcommand{\rv}[1]{{\bf{#1}}}

\newcommand{\leak}{\tn{leak}_{\tn{EC}}}

\newcommand{\ignore}[1]{} 

\begin{document}
\title{Finite-key analysis for time-energy high-dimensional quantum key distribution}

\author{Murphy Yuezhen Niu}
\affiliation{Research Laboratory of Electronics, Massachusetts Institute of Technology, 77 Massachusetts Avenue, Cambridge, Massachusetts 02139, USA}
\affiliation{Department of Physics, Massachusetts Institute of Technology, Cambridge, Massachusetts 02139, USA}
\author{Feihu Xu}
\affiliation{Research Laboratory of Electronics, Massachusetts Institute of Technology, 77 Massachusetts Avenue, Cambridge, Massachusetts 02139, USA}

\author{Fabian Furrer}
\affiliation{NTT Basic Research Laboratories, NTT Corporation,
3-1 Morinosato-Wakamiya, Atsugi, Kanagawa, 243-0198, Japan}
\author{ Jeffrey H. Shapiro}
\affiliation{Research Laboratory of Electronics, Massachusetts Institute of Technology, 77 Massachusetts Avenue, Cambridge, Massachusetts 02139, USA}

\date{\today}
\begin{abstract}

Time-energy high-dimensional quantum key distribution (HD-QKD) leverages the high-dimensional nature of time-energy entangled biphotons and the loss tolerance of single-photon detection to achieve long-distance key distribution with high photon information efficiency. To date, the general-attack security of HD-QKD has only been proven in the asymptotic regime, while HD-QKD's finite-key security has only been established for a limited set of attacks. Here we fill this gap by providing a rigorous HD-QKD security proof for general attacks in the finite-key regime. Our proof relies on a novel entropic uncertainty relation that we derive for time and conjugate-time measurements using dispersive optics, and our analysis includes an efficient decoy-state protocol in its parameter estimation.  We present numerically-evaluated secret-key rates illustrating the feasibility of secure and composable HD-QKD over metropolitan-area distances when the system is subjected to the most powerful eavesdropping attack.

\end{abstract}
\maketitle


\section{Introduction}

Quantum key distribution~(QKD) enables secure communication based on fundamental laws of quantum physics~\cite{bennett1984quantum,ekert1991quantum}, as opposed to the security that is presumed from computational complexity in conventional public-key cryptography. Current work on QKD focuses on patching security holes in practical implementations, increasing secret-key rates and secure-transmission distances, and unifying understanding of the many different protocols~\cite{lo2014review}. Existing QKD protocols can be divided into two major categories: discrete-variable (DV)~\cite{bennett1984quantum, Huang2003, Wang2003, Lo2005} and continuous-variable (CV)~\cite{grosshans2002continuous} QKD. The predominant DV-QKD is more robust to loss than CV-QKD, and thus offers longer secure-transmission distance~\cite{liu2010decoy,wang2012,korzh2015, curty2014finite}. CV-QKD, on the other hand, offers higher photon information efficiency (PIE) than DV-QKD, and thus potentially higher key rates at short distances~\cite{jouguet2013experimental}.

High-dimensional QKD (HD-QKD) exploits the best features of DV and CV protocols to simultaneously achieve high PIE and long secure-transmission distance~\cite{cerf2002security,Gisin2004,WalmsleyPRL2008,lee2014entanglement,zhong2015photon,mirhosseini2015high, Boyd2015}. One of the most appealing candidates for implementation is time-energy HD-QKD~\cite{zhong2015photon,Kwiat2013,mower2013high,lee2013finite, zhang2014unconditional,DariusPRA2015, bao2016finite}. It generates keys using the detection times of time-energy entangled photon pairs, whose continuous nature permits encoding of extremely large alphabets. The security analysis of time-energy HD-QKD has been improving ever since the protocol was proposed~\cite{Kwiat2013,mower2013high,DariusPRA2015,lee2013finite,zhang2014unconditional,bao2016finite}. Nevertheless, a rigorous security proof that satisfies the composability condition~\cite{renner2005security} and takes full account of the finite-size effects against \emph{general} attacks (the most powerful eavesdropping attack) has been missing. For this reason, the feasibility of secure, metropolitan-area, time-energy  HD-QKD using a reasonable time interval for signal transmission has yet to be fully established.

In this paper we make three contributions.  First, we derive a new entropic uncertainty relation between time and conjugate-time measurements that are made via non-local dispersion cancellation. Second, we use the new uncertainty principle to prove the composable security of time-energy HD-QKD  in the finite-key regime against general (coherent) attacks. Third, we find the dispersion strength for the conjugate-time basis transformation~\cite{mower2013high} that maximizes HD-QKD's secret-key rate.

The entropic uncertainty relation is indispensable for analyzing general attacks against time-energy HD-QKD. Although an entropic uncertainty relation for field quadratures has been developed~\cite{furrer2014position}, and applied recently to CV-QKD security analysis~\cite{furrer2012continuous}, it cannot be directly applied to time-energy HD-QKD because time and conjugate-time measurements are not described by maximally incompatible operators \cite{walach2001}, such as position and momentum.  To overcome this challenge, we construct a new entropic uncertainty relation specifically for time and conjugate-time measurements. Because entropic uncertainty relations figure prominently in quantum metrology~\cite{metrology}, quantum randomness certification~\cite{random,XuQRNG}, entanglement witnesses~\cite{entanglement1, entanglement2}, two-party cryptography~\cite{crypto1, crypto2}, QKD security analysis~\citep{curty2014finite,QKD,tomamichel2011uncertainty,tomamichel2012tight, Wang2013, Zhou2014}, and other applications~\cite{Patrick2015}, we expect that our uncertainty relation for time and conjugate-time measurements may have uses well beyond what will be presented below.

The secret-key rate formula we obtain using our entropic uncertainty relation allows us to verify important advantages that HD-QKD offers over alternative protocols.  In particular, HD-QKD offers higher PIE~(3.3 bits/photon) than both CV-QKD~(0.5 bits/photon~\cite{furrer2014reverse}) and DV-QKD~(0.1 bits/photon~\cite{lim2014concise}), thus ensuring higher secret-key rates under photon-starved conditions, in which the photon-detection rate is much lower than the photon-generation rate because of the loss incurred in long-distance propagation and the relatively long recovery times of available single-photon detectors.  Also, HD-QKD offers a longer maximum secure-transmission distance for general attacks (e.g., 160\,km for a 30-min session using the system parameters given below in Table~\ref{Tab:exp:parameters}) as compared to that for CV-QKD~\cite{furrer2012continuous,leverrier2013security}, even in the case of reverse reconciliation (e.g., 16\,km~\cite{furrer2014reverse}).  Furthermore, because our entropic uncertainty relation is parametrized by the HD-QKD protocol's time-bin duration, $\delta$, and conjugate-time basis transformation's group-velocity dispersion (GVD) coefficient, $\beta_D$, optimizing the $\beta_D$ value can increase HD-QKD's secure-transmission distance to 210\,km---and provide a 17\,Mbit/s expected secret-key rate at zero distance---without resorting to a higher clock rate.

The remainder of the paper is organized as follows. The HD-QKD protocol is described briefly in Sec.~\ref{secProtocol}, with a detailed account---including its use of decoy states for channel estimation---appearing in Appendix~\ref{protocalAPP}.  The security analysis for coherent attacks in the finite-key regime is contained in Sec.~\ref{secSecureAnalysis}.  Its security proof relies on the entropic uncertainty relation that is derived in Sec.~\ref{secUncertainty}. (For comparison, the entropic uncertainty relation obtained from the conventional dilation assumption is presented in Appendix~\ref{InfiniteOverlap}.)  A numerical evaluation of HD-QKD's secret-key rate and PIE follows in Sec.~\ref{secNumerics}, which illustrates the advantages offered by this protocol, and Sec.~\ref{Conclusion} provides summarizing discussion.

\section{Protocol}\label{secProtocol}
Time-energy HD-QKD that relies on dispersive optics works as follows~\cite{mower2013high,lee2013finite}.  In each round, Alice generates a time-energy entangled photon pair from a spontaneous parametric down-conversion~(SPDC) source, sends one photon to Bob and retains the other.  Alice and Bob choose independently and at random to measure their photons in either the time basis ($\mathsf{T} $) or the conjugate-time basis ($\mathsf{W}$), where the latter is a dispersive-optics proxy for a frequency measurement. Alice and Bob discretize their outcomes into time bins of duration $\delta$. The process repeats for $N$ rounds until Alice and Bob obtain enough detections to begin post-processing. At the end of all measurements, the two sides reveal their basis choices and discard all data measured using mismatched bases. Secret keys are extracted from the events in which Alice and Bob both chose the $\mathsf{T}$ basis, while the $\mathsf{W}$ basis outcomes are publicly announced for parameter estimation. Using the decoy-state method~\cite{ Huang2003, Wang2003, Lo2005, DariusPRA2015,lim2014concise}, Alice and Bob estimate the number of detections in $\mathsf{T}$ that were generated from single-pair SPDC emissions, and the corresponding $\text{L}_{1}$ code distance in the $\mathsf{W}$ basis, see Appendix~\ref{App:decoy} for the details. They abort the protocol if this distance exceeds a predetermined value $d_0$ (see Appendix~\ref{thresholdmodeling}). Otherwise, they perform error correction and privacy amplification to generate the secret key.

The conjugate-time measurement for the $\mathsf{W}$ basis is realized by direct detection at Alice and Bob's terminals after they have sent their photons through normal and anomalous GVD elements, respectively~\cite{mower2013high,lee2013finite}. These GVD elements' dispersion coefficients have equal magnitudes (and opposite signs) so their effects are non-locally canceled~\cite{Franson1992}.  As a result, Alice and Bob's $\mathsf{W}$-basis measurements are as strongly correlated as those in the $\mathsf{T}$ basis, i.e., the dispersion transformation allows them to perform a spectral-correlation measurement with only time-resolved single-photon detection~\cite{mower2013high,lee2013finite}.

\section{Security Analysis}\label{secSecureAnalysis}

\subsection{Security Definition}\label{secDef}
Given that the parameter-estimation test is passed with probability $p_{\text{pass}}$,  Alice and Bob end up with final keys that are classical random vectors, $\rv{K}_{\rm A}$ and $\rv{K}_{\rm B}$, which might be correlated with a quantum system,  $\rv{E}$, held by Eve. Mathematically, this situation corresponds to a classical-quantum state $ \rho_{\rm K_AE}=\frac{1}{|\mathcal{S}|}\sum_{s}\proj{s}\otimes \rho_{\rm E}^s  $, where $\{\ket{s}\}$ denotes an orthonormal basis for Alice's dimension-$|\mathcal{S}|$ key space, and the subscript ${\rm E}$ indicates Eve's quantum state. We characterize a QKD protocol by its correctness and secrecy. For that we use a notion of security based on the approach developed in~\cite{renner2005security}. A protocol is called $\epsilon_{c}$\emph{-correct} if the probability that $\rv{K}_{\rm A}$ differs from $\rv{K}_{\rm B}$ is smaller than $\epsilon_c$. We say that a protocol is $\epsilon_s$\emph{-secret} if the state $\rho_{\rm K_AE}$ is $\epsilon_s$-close to the ideal situation described by the tensor product of uniformly distributed keys on Alice's side and Eve's quantum state, $U_{\rm K_A} \otimes \rho_{\rm E}$, such that $p_{\text{pass}}\|\rho_{\rm K_AE}-U_{\rm K_A}\otimes\rho_{\rm E}\|_1 \leq \epsilon_s$. A QKD protocol is then said to be $\epsilon$-secure if it is both $\epsilon_c$-correct and $\epsilon_s$-secret, with $\epsilon_c+\epsilon_s\leq\epsilon$. Our security definitions ensure that the protocol remains secure in combination with any other protocol, i.e., the protocol is secure in the universally composable framework~\cite{renner2005security}.

\subsection{Assumptions}

Before deriving our lower bound on secret-key length, we first specify the assumptions that will be employed: (1) Alice's SPDC source produces independent, identically-distributed biphotons whose correlation time and coherence time are well characterized. (2) For each pump pulse, Alice is able to randomly set her SPDC source's biphoton intensity (mean photon-pairs generated per pump pulse) to be either $\mu_1$, $\mu_2$, or $\mu_3$ with probabilities $p_{\mu_1}$, $p_{\mu_2}$, and  $p_{\mu_3}$. (3). Alice and Bob's laboratories are secure, i.e., free from any information leakage. (4) Alice and Bob independently and randomly choose between measuring in the time and conjugate-time bases with probabilities $q$ and $1-q$.  Most of these assumptions are already made in conventional CV-QKD and DV-QKD security analysis.

\subsection{Security Proof}\label{secProof}

In order to characterize information leakage in a realistic quantum communication system with a finite number of communication rounds, we use smooth min-entropy instead of von Neumann entropy~\cite{renner2005security,tomamichel2011leftover}. Discretizing Alice and Bob's photon-detection times to time bins of duration $\delta$ results in data vectors comprised of integers representing bin numbers.  In particular, with random vectors $\rv{X}_{\rm A}$ and $\rv{X}_{\rm B}$ denoting  Alice and Bob's raw keys from her $\mu_1$-intensity transmissions, Eve's uncertainty (lack of knowledge) is measured by her difficulty in guessing Alice's raw key $\rv{X}_{\rm A}$, i.e., the conditional smooth min-entropy $H_{\min}(\rv{X}_{\rm A}|\rv{E})$, where $\rv{E}$ denotes Eve's quantum state. $H_{\min}(\rv{X}_{\rm A}|\rv{E})$ quantifies the randomness that can be extracted from $\rv{X}_{\rm A}$ which is statistically independent of $\rv{E}$~\cite{renner2005security,tomamichel2011leftover} with error probability $\epsilon$.

The secret-key length $\ell$ that is $\epsilon_{s}$-secret is given by~\cite{renner2005security}
\begin{align}\label{Eq:GeneralKeyLength}
  \ell \geq H_{\min}^{\epsilon}(\rv{X}_{\rm A}|\rv{E}) - \leak + \log_2(\epsilon_s^2\epsilon_c).
\end{align}
Here, $\leak$ is the information leaked to Eve during error correction, which can be directly measured during that correction process, and
$H_{\min}^{\epsilon}(\rv{X}_{\rm A}|\rv{E})$ is the smooth min-entropy maximized over states that are
$\epsilon$ close to the classical-quantum state $\rho_{\rm X_AE}=\frac{1}{|\mathcal{S}|}\sum_{s}\proj{s}\otimes \rho_{\rm E}^s$. The correctness of the protocol is guaranteed by the key-verification step, which uses a two-universal hash function to ensure that Bob's corrected key differs from Alice's with probability at most $\eps_\tn{hash}$, implying that the protocol is $\eps_{c}$-correct with $\eps_c=\eps_\tn{hash}$.

The essential insight is that Eve's information about the $\mu_1$-intensity, $\mathsf{T}$-basis detection times can be bounded using the complementary $\mathsf{W}$-basis measurements. In particular, if Alice and Bob's $\mathsf{W}$-basis measurements are highly correlated, then Eve's knowledge about the outcome of their $\mathsf{T}$-basis measurements is nearly zero, because the two observables are \emph{incompatible}.

Let $\rv{Y}_{\rm A}$  and $\rv{Y}_{\rm B}$ be Alice and Bob's random vectors of $\mu_1$-intensity conjugate-time measurement outcomes.  Without loss of generality we set the length of these four classical strings to be equal: $ |\rv{X}_{\rm A}|=|\rv{X}_{\rm B}|=|\rv{Y}_{\rm A}|=|\rv{Y}_{\rm B}|= n_{\sT,\mu_1}$. Then, from~\cite{tomamichel2012tight,furrer2012continuous,tomamichel2011uncertainty,furrer2014position}, we have the uncertainty relation:
\begin{eqnarray}\label{eq:UR1}
H_{\min}^{\epsilon}(\rv{X}_{\rm A}| \rv{E} )+H_{\max}^{\epsilon }(\rv{Y}_{\rm A}| \rv{Y}_{\rm B} )\geq -n_{\sT,\mu_1} \log_2[c(\delta, \beta_D)],
\end{eqnarray}
where the smooth max-entropy $H_{\max}^{\epsilon }(\rv{Y}_{\rm A} | \rv{Y}_{\rm B} )$ measures the amount of information needed to reconstruct $\rv{Y}_{\rm A} $ given $\rv{Y}_{\rm B}$ with error probability bounded above by $\epsilon$, and $c(\delta, \beta_D)$ is the overlap between the time and conjugate-time measurement operators, which depends on $\delta$, the time-bin duration, and $\beta_D$, the magnitude of the GVD elements' dispersion coefficient. 

With $\boldsymbol{\Pi}=\{\Pi_n\}$ and $\boldsymbol{\Pi'} = \{\Pi'_m\}$ being an arbitrary pair of positive operator-valued meausurements (POVMs), their overlap, $c(\boldsymbol{\Pi},\boldsymbol{\Pi'}) = \sup_{n,m} \left\Vert \sqrt{\Pi_n} \sqrt{\Pi'_m}\right\Vert^2$, quantifies their incompatibility, i.e., lower values of $c(\boldsymbol{\Pi},\boldsymbol{\Pi'})$ mean increased incompatibility.  Our uncertainty bound involves the overlap-quantified incompatibility between the time and conjugate-time POVMs whose outcomes are used for key generation and parameter estimation, respectively. Typically, see Sec.~\ref{secNumerics}, lower $c(\delta,\beta_D)$ values allow longer secret keys to be extracted.  Our tri-partite entropic uncertainty relation and the security analysis that follows therefrom are adapted from CV-QKD's finite-key analysis~\cite{furrer2012continuous}, an approach that works for all QKD protocols which rely on a pair of incompatible continuous measurements for key generation and parameter estimation.  In our case, the security analysis requires accounting for our use of  discretized time and conjugate-time measurements that are obtained from underlying continuous POVMs.  Note that the different measurement operators employed in different QKD protocols lead to different overlap behaviors in their entropic uncertainty relations.

The major difficulty in determining $c(\delta, \beta_D)$ for our protocol comes from the absence of negative energy for electromagnetic-field modes, which implies that under the conventional commutation relation, the time-measurement operator cannot be projective~\cite{busch1994,delgado1997}, thus preventing existing results~\cite{Fabian2015} being applied to the time and conjugate-time POVMs. We can, however, dilate the time and conjugate-time operators by forsaking the constraint of positive frequency on photon-annihilation operators~\cite{werner1987,kiukas2012}.  Such dilations are well justified for the quantum theory of coincidence measurement~\cite{Jeff,Franson1992}, because the negative frequency components do not contribute to detection outcomes.  But, because we are not assured that the dilation-assumption $c(\delta,\beta_D)$ will suffice for our security proof, we derive the following entropic uncertainty relation for time and conjugate-time measurements \emph{without} dilation in Sec.~\ref{secUncertainty}:
\begin{align}\label{newUncertainty}
H_{\min}^{\epsilon}(\rv{X}_{\rm A}| \rv{E} )+H_{\max}^{\epsilon }(\rv{Y}_{\rm A}| \rv{Y}_{\rm B} )\geq -n_{\sT,\mu_1} \log_2\!\left(\frac{1.37\delta^2}{2\pi^2 \beta_D}\right),
\end{align}

Next, we use a generalized chain-rule result~\cite{vitanov2013chain} to decompose $\rv{X}_{\rm A}$ into $\rvX^\tn{0}_{\rm A}\rvX^{\tn{1}}_{\rm A}\rvX^{\tn{m}}_{\rm A}$, which is a concatenation of the raw keys arising from vacuum, single-pair, and multi-pair coincidences. Neglecting the multi-pair contribution, we have $n_{\sT,\mu_1} \geq  (\underline{n_{\sT,0}}+ \underline{n_{\sT,1}})$, with  $\underline{n_{\sT,0}}$  and $\underline{n_{\sT,1}}$ being lower bounds on $n_{\sT,0}$ and $n_{\sT,1}$, the coincidence-count contributions from vacuum and single-pair events, respectively, when Alice's SPDC intensity is  $\mu_1$ (see Appendix~\ref{App:decoy}). We then have the following lower bound on the smooth min-entropy~\cite{lim2014concise}:
\begin{align}
  H_{\min}^{\epsilon}(\rv{X}_{\rm A}|\rv{E}) &\geq  -(\underline{n_{\sT,0}}+ \underline{n_{\sT,1}})\log_2[c(\delta, \beta_D)] \nonumber \\
   & \hspace*{.1in}- H_{\max}^{\epsilon}(\rv{Y}_{\rm A}|\rv{Y}_{\rm B})\ .\label{Eq:uncertainty}
\end{align}
Using a result from CV-QKD~\cite{furrer2012continuous}, we get the following upper bound on the smooth max-entropy:
\begin{align}\label{Eq:Hmax}
H_{\max}^{\epsilon}(\rv{Y}_{\rm A}|\rv{Y}_{\rm B}) \leq  n_{\sT,\mu_1}\log_2[\gamma(d_0+ \Delta)],
\end{align}
where  $\gamma(x)$ obeys
\begin{align}\label{Eq:gamma}
\gamma(x)= \left(x+\sqrt{1+x^2}\right)\!\Big(\frac{x}{\sqrt{1+x^2}-1}\Big)^{x}.
\end{align}
The $\Delta$ parameter is the statistical fluctuation that quantifies how well the data subset used for parameter estimation represents the entire dataset,
\begin{align}\label{Eq:Delta}
\Delta \approx \frac{T_f}{\delta} \sqrt{\frac{1}{q^2(1-q)^2\underline{n_{\sT,01}}}\ln\frac{1}{\epsilon_s/4-2f(p_{\alpha},\underline{n_{\sT,01}})}},
\end{align}
where $f(p_{\alpha},\underline{n_{\sT,01}})=\sqrt{2(1-(1-p_{\alpha})^{\underline{n_{\sT,01}}})}$. and $p_{\alpha}$ is the probability, for a given pump pulse, that Alice and Bob detect photons separated by more than a frame duration, $T_f$, and $\underline{n_{\sT,01}}= \underline{n_{\sT,0}}+ \underline{n_{\sT,1}}$.

Combining the preceding results, we obtain the following lower bound on the secret-key length:
\begin{align}
\ell  & \ge -\underline{n_{\sT,01}}\log_2[c(\delta, \beta_D)]-  n_{\sT,\mu_1}\log_2[\gamma(d_0+ \Delta)] \nonumber \\
& \hspace*{.2in}- {\rm leak}_{\text{EC} } + \log_2(\epsilon_s^2\epsilon_c).\label{securekey}
\end{align}

\section{Time-Conjugate Time Entropic Uncertainty Relation}\label{secUncertainty}
To justify \eqref{newUncertainty}, we only need to evaluate the overlap, $c(\delta,\beta_D)$, in \eqref{eq:UR1} for the discretized single-photon time and conjugate-time measurement operators that derive from their continuous-time counterparts, $T(t)$ and $W(t)$, by coarse-graining to time bins of duration $\delta$. Here, we omit polarization degrees of freedom as they do not affect the overlap. Our starting point is the infinite-dimensional version of the general uncertainty relation for smooth min-entropy and smooth max-entropy~\cite{tomamichel2011uncertainty} that was derived in~\cite{furrer2014position}.

We use $\ket{\omega}=a^{\dagger}(\omega+\omega_0)\ket{0}$ to denote the single-photon state detuned by frequency $\omega$ from some fixed center frequency $\omega_0$. (Later, this center frequency will be $\omega_P/2$, i.e., half the SPDC source's pump frequency.)  This state satisfies the orthonormality condition $\braket{\omega_1}{\omega_2} = 2\pi\,\delta(\omega_1 - \omega_2)$. The single-photon Hilbert space is simply $\cH = L^2(\Omega) $, i.e., the space of square-integrable, complex-valued functions on the frequency-domain region $\omega \in \Omega \equiv  [\omega_{\rm min},\infty)$, where the minimum detuning satisfies $\omega_{\rm min} \ge \omega_0$. In particular, we associate a function $f\in L^2(\Omega)$ to the state
\begin{equation}
\ket{f} = \int_\Omega\ddo{\omega} f(\omega)\ket{\omega}\, ,
\end{equation}
so the inner product between two such states, $\ket{f}$ and $\ket{g}$, is $\braket{f}{g} = \int_\Omega \ddo{\omega} f^*(\omega)g(\omega)$.

Using the above notation we have that the time-measurement operator $T(t)$ can be expressed as
\begin{equation} \label{eq:Tobs}
T(t) = \int_\Omega \ddo{\omega_1}\int_\Omega \ddo{\omega_2} e^{i(\omega _1-\omega_2)t} \ketbra{\omega_1}{\omega_2} = \ketbra{\phi_t}{\phi_t} \, ,
\end{equation}
where $\phi_t(\omega)=e^{i\omega t}$. Similarly, we can write
\begin{align} \label{eq:Wobs}
W(t) &= \int_\Omega \ddo{\omega_1}\int_\Omega\ddo{\omega_2} e^{i(\omega_1 -\omega_2)t}e^{i\beta_D(\omega_1^2 -\omega_2^2)/4} \ketbra{\omega_1}{\omega_2}\\\nonumber
& = \ketbra{\psi_t}{\psi_t} \, ,
\end{align}
where $\psi_t(\omega)=e^{i(\omega t + \beta_D\omega^2/4)}$. We then introduce partitions, $\{I_k\}$ and $\{J_k\}$, of the time and conjugate-time axes, from which we obtain the coarse-grained versions of $T(t)$ and $W(t)$, namely the POVMs $T^\delta = \{T_k\}$ and $W^\delta = \{W_k\}$, where
\begin{align}\label{eq:CoarseGrained}
T_k= \int_{I_k}\!{\rm d}t\,  T(t) \quad \text{and} \quad W_k= \int_{J_k}\!{\rm d}t\, W(t)\,.
\end{align}
From \cite{tomamichel2011uncertainty, furrer2014position} the overlap for these discrete POVMs satisfies
\begin{align}
c(\delta,\beta_D)& =\bar{c}(T^\delta ,W^\delta ) \nonumber \\
&= \sup_{k,l} \left\Vert \sqrt{T_k} \sqrt{W_l}\right\Vert^2 \nonumber \\
&=\sup_{s,t}\left\Vert \sqrt{ T^\delta(s)} \sqrt{W^\delta(t)} \right\Vert^2,\label{eq:Overlap1}
\end{align}
where $T^{\delta} (s)= \int_{s}^{s+\delta}\!{\rm d}u\, T(u)$ and  $W^{\delta}(t) = \int_{t}^{t+\delta}\!{\rm d}u\, W(u)$.

Because the $\{T_k\}$ and $\{W_k\}$ are not projective, it is difficult to evaluate Eq.~\eqref{eq:Overlap1} directly. Instead, we will use the approximation from~\cite{furrer2014position}, in which an uncertainty relation is derived in the continuous-time case. We take $ \cT $ and $ \cW$ to represent the continuous-time classical outcomes of the time and conjugate-time measurements, and $ \cT_\delta $ and $ \cW_\delta$ to be their discretized versions. From~\cite{furrer2014position} we have that
\begin{align} \label{eq:diffboundMin}
H_{\min}(\cT_\delta|\rv{E}) & \geq h_{\min}(\cT|\rv{E}) - \log_2(\delta) \\[.05in]
H_{\max}(\cW_\delta|\rv{B}) & \geq h_{\max}(\cW|\rv{B}) - \log_2(\delta) \,, \label{eq:diffboundMax}
\end{align}
where $h_{\min}(\cT|\rv{E})$ and $h_{\max}(\cT|\rv{B})$ are the differential min-entropy and differential max-entropy of the continuous-time outcome $\cT$ conditioned on Eve's state ($\rv{E}$) and Bob's state ($\rv{B}$), respectively.  We also know that these differential entropies satisfy~\cite{furrer2014position}
\begin{equation}
h_{\min}(\cT|\rv{E}) + h_{\max}(\cW|\rv{B})  \geq -\log_2[\bar{c}_\infty(T,W)] \, ,
\end{equation}
where
\begin{equation}\label{eq:Overlap3}
\bar{c}_\infty (T,W) =\liminf_{\delta \rightarrow 0}\!\left[ \frac{\bar{c}(T^\delta,W^\delta)}{\delta^2}\right] \, .
\end{equation}
Inequalities~\eqref{eq:diffboundMin} and~\eqref{eq:diffboundMax} yield the following uncertainty relation for coarse-grained measurements:
\begin{align}\label{Hmindelta}
H_{\min}(\cT_\delta|\rv{E}) + H_{\max}(\cW_\delta|\rv{B})  \geq  -\log_2[\bar{c}_\infty(T,W) \delta^2].
\end{align}

We can find the overlap for the differential entropies via
\begin{align}
\bar{c}_\infty (T,W) &= \liminf_{\delta \rightarrow 0}\!\left[\frac{\bar{c}(T^\delta,W^\delta)}{\delta^2}\right] \nonumber \\
&=\sup_{s,t}\liminf_{\delta \rightarrow 0}\!\left[\frac{1}{\delta^2} \left\Vert\sqrt{T^\delta(s)} \sqrt{W^\delta(t)}  \right\Vert^2\right] \nonumber \\
 &  =\sup_{s,t}\liminf_{\delta \rightarrow 0}\left\Vert \sqrt{\frac{T^\delta(s)}{\delta} } \sqrt{\frac{W^\delta(t)}{\delta}} \right\Vert^2  \nonumber \\
& =\sup_{s,t}\left\Vert \sqrt{T(s)}\sqrt{ W(t)}\right\Vert ^2 \, ,
\end{align}
where we have used $\lim_{\delta \to 0}\frac{1}{\delta} \int_{s}^{s+\delta}\!{\rm d}t\, T(t) = T(s)$ and similarly for $W$. Inserting the definitions of $T(s)$ and $W(t)$ from Eqs.~\eqref{eq:Tobs} and \eqref{eq:Wobs}, we obtain
\begin{equation} \label{eq:cinf}
\bar{c}_\infty(T,W) = \sup_{s,t} | \braket{\phi_s}{\psi_t}|^2 \, .
\end{equation}
A simple calculation now gives us
\begin{align}
 \bar{c}_\infty(T,W)  = \sup_{s,t} \, \left|  \int_\Omega\!\frac{{\rm d}\omega}{2\pi}\, e^{i\omega(t-s)}e^{ -i\beta_D\omega^2/4} \right|^2 \, ,
\end{align}
For $\Omega = [\omega_{\min},\infty)$, performing the optimization with $\omega_{\min} > -\omega_0$ and $-\infty < t,s < \infty$ yields the maximum overlap
\begin{equation}\label{finitecdelta}
\bar{c}_\infty(T,W) \approx \frac{1.37}{2\pi^2 \beta_D} \, .
\end{equation}
Inserting the above result into~(\ref{Hmindelta}) gives us the overlap for the discrete measurements used in the secret-key length bound
\begin{align}\label{finaloverlap}
c(\delta,\beta_D) \approx \frac{1.37\delta^2}{2\pi^2 \beta_D}.
\end{align}
This uncertainty bound is \emph{tighter} than the $c(\delta, \beta_D) =\delta^2/2\pi^2 \beta_D$ overlap, obtained in Appendix~\ref{InfiniteOverlap}, when dilation is used by taking $\Omega=(-\infty, \infty)$ so that the $T(t)$ and $W(t)$ operators become projective and maximally incompatible, i.e., analogous to position and momentum.   These overlap results showcase the subtle difference between the entropic uncertainty relation of quantum time and conjugate-time measurements and that of the homodyne measurements from~\cite{furrer2012continuous}.  Indeed, the factor of 1.37 in Eq.~\eqref{finaloverlap} is crucial for the general-attack security of HD-QKD, because a secret-key length that presumed the dilation result for the overlap would be insecure.

\section{Performance Example}\label{secNumerics}

\begin{table}[hbt]
\centering
\begin{tabular}{c @{\hspace{0.3cm}} c @{\hspace{0.3cm}} c @{\hspace{0.3cm}} c @{\hspace{0.3cm}} c @{\hspace{0.3cm}} c} \hline \hline
$\eta_{d}$ & $Y_{0}$ & $\sigma_{\rm jit}$ & $\alpha$ &  $\beta_D$ & $R_{\rm rep}$  \\
90\% & 1\,kHz & 18\,ps & 0.21\,dB/km &  $2\times 10^4$\,ps$^2$ & 55.6\,MHz~\cite{Footnote1}\\
\hline
$\sigma_{\rm cor}$ & $\sigma_{\rm coh}$ & $\delta$ & $\beta_e$ & $q$ & $\epsilon$ \\
2\,ps & 6\,ns & 20\,ps & 0.91 & 0.9 & $10^{-10}$ \\
\hline \hline
\end{tabular}
\caption{List of parameters, mostly from~\cite{zhong2015photon}, used in numerical evaluation: detection efficiency $\eta_{d}$, dark-count rate $Y_{0}$, detector time jitter $\sigma_{\rm jit}$~\cite{Tang2012}, fiber-loss coefficient $\alpha$, GVD coefficient $\beta_D$, system clock rate $R_{\rm rep}$, biphoton correlation time $\sigma_{\rm cor}$, pump coherence time $\sigma_{\rm coh} \approx T_f $, time-bin duration $\delta$, reconciliation (error-correction) efficiency $\beta_e$, probability of choosing the time basis $q$, and overall security bound $\epsilon$.} \label{Tab:exp:parameters}
\end{table}

\begin{table}[hbt]
\centering
\begin{tabular}{c @{\hspace{0.3cm}} c @{\hspace{0.3cm}} c @{\hspace{0.3cm}} c } \hline \hline
Parameters & BB84~\cite{lim2014concise} & CV-QKD~\cite{furrer2014reverse} & HD-QKD\\ \hline
PIE (bits/photon)\footnote{PIE in HD-QKD is defined as secret bits per single photon detection by Bob given that Alice has made a detection in the same basis; PIE in BB84 is defined as secret bits per use~\cite{lucamarini2013efficient}; and PIE in CV-QKD is defined as secret bits per signal~\cite{furrer2014reverse}.}& $\approx$ 0.1 & 0.5 &  3.3 \\
Key rate at 0 Dist (bits/s) & $\approx$ 8\,M\footnote{Assumes a decoy-state BB84 system with a 1\,GHz clock rate~\cite{lucamarini2013efficient}.} & $\approx$ 6\,M\footnote{Assumes a CV-QKD system with the same 55.6\,MHz clock rate as HD-QKD.} & 8.6\,M \\
Max Dist. (km) & 170 & 16 & 96\\
\hline \hline
\end{tabular}
\caption{Performance comparison for different protocols with finite-key analysis against general attacks. The first and second rows compare the PIEs and the secret keys rate at 0 km fiber length. The third row compares the maximum secure-transmission distance. All three protocols are evaluated at the a block size of $ 10^9$, equivalent to a 1\,min running time in HD-QKD with parameters specified in Table~\ref{Tab:exp:parameters}.} \label{Tab:parameter}
\end{table}

\begin{figure*}
  \includegraphics[width=7in]{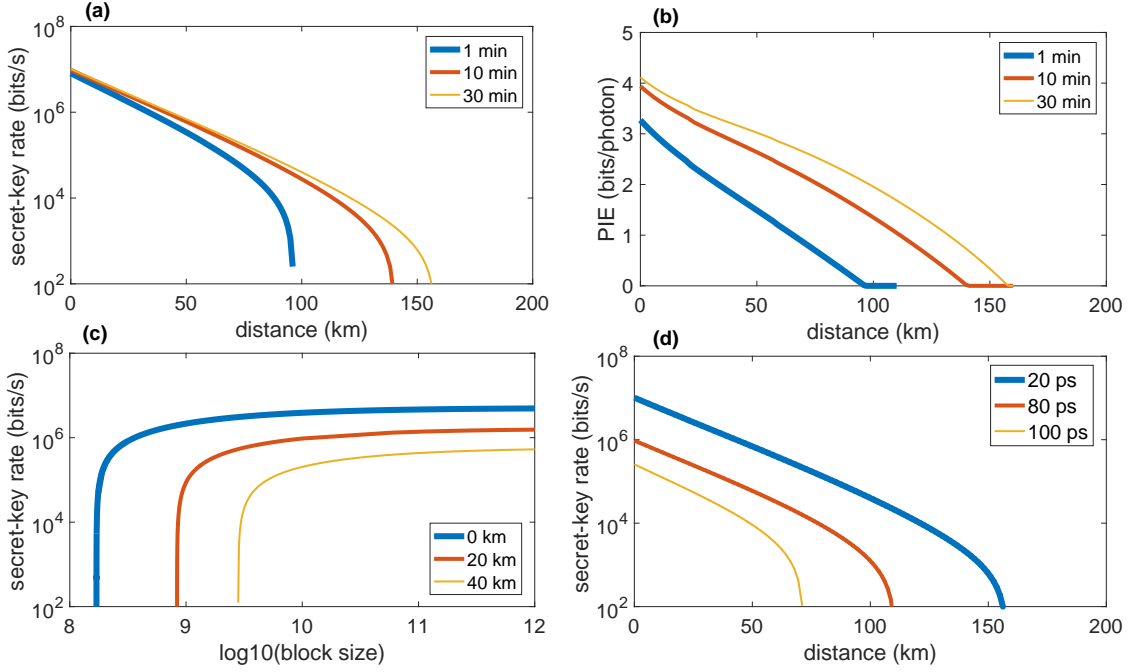}\\
  \caption{Numerically-evaluated performance of time-energy HD-QKD with threshold code distance $d_0 =2$ and other parameters as listed in Table~\ref{Tab:exp:parameters}. (a) Secret-key rate~(bits/s) versus transmission distance~(km) for different total running times of transmission: top curve (yellow) 30\,min, middle curve (red) 10\,min, bottom curve (blue) 1\,min. (b) PIE (bits/photon) versus transmission distance~(km) for different running times:  top curve (yellow) 30\,min, middle curve (red) 10\,min, bottom curve (blue) 1\,min. (c) Secret-key rate~(bits/s) versus block size (running time/clock rate) for different transmission distances:  top curve (blue) 0\,km, middle curve (red) 20\,km, bottom curve (yellow) 40\,km. (d) Secret-key rate~(bits/s) versus transmission distance~(km) for different time-bin durations $\delta$, where the running time is fixed at 30\,min:  top curve (blue) 20\,ps, middle curve (red), 80\,ps, bottom curve (yellow) 100\,ps.}\label{allfigure}
\end{figure*}

Based on the secret-key rate formula~(\ref{securekey}), we  numerically evaluated the performance of the time-energy HD-QKD protocol in the finite-key regime under general attacks. See Table~\ref{Tab:exp:parameters} for the parameters that were assumed. The calculated secret-key rates and PIEs at different lengths of standard telecom fiber are shown in Figs.~\ref{allfigure}(a) and~\ref{allfigure}(b). We see that HD-QKD can easily tolerate a 100\,km standard fiber within a reasonable running time for transmission (e.g., 10\,min). This secure-transmission distance significantly exceeds that of CV-QKD (around 10\,km~\cite{furrer2014reverse}). In addition, the secret-key rate of HD-QKD at zero distance is about 8.6\,Mbit/s (see Table~\ref{Tab:parameter}), which is comparable to that of CV-QKD with the same 55.6\,MHz clock rate, and to that of decoy-state BB84 with a state-of-the-art 1\,GHz clock rate~\cite{lucamarini2013efficient}. Moreover, HD-QKD can offer a higher PIE, up to 4.3\,bits/photon (with 30\,min running time), than does decoy-state BB84~\cite{lim2014concise}, whose PIE can never exceed 1\,bit per use.

In Fig.~\ref{allfigure}(c) we show the secret-key rate as a function of block size.  Here we see that the minimum required block size for HD-QKD is slightly larger than those of decoy-state BB84~\cite{lim2014concise} and CV-QKD~\cite{furrer2014reverse}. Finally, Fig.~\ref{allfigure}(d) plots the secret-key rate versus transmission distance for different time-bin durations, showing that shorter duration time bins offer higher key rates for a given biphoton source. We remark that detectors with less than 20\,ps jitter have already been demonstrated in recent experiments~\cite{Tang2012}.

\begin{figure}[h]
  \includegraphics[width = 3.42in]{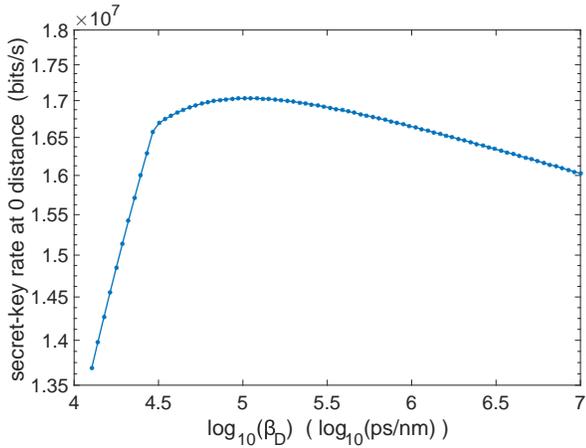}\\
  \caption{Secret-key rate at zero distance versus GVD coefficient $\log_{10}(\beta_D)$ for 30\,min running time. The conventional units for $\beta_D$ are employed here: ps per nm of bandwidth at telecom wavelength.  The secret-key rate at zero distance achieves its 17\,Mbit/s maximum at  $\beta_D = 10^5$\,ps/nm at telecom wavelength, which is equivalent to $2\times 10^6$\,ps$^2$ in the units used in Table~\ref{Tab:exp:parameters}.}\label{dispersion}
\end{figure}

Our work clarifies how the secret-key rate of time-energy HD-QKD using dispersive-optics depends on the time-bin duration $\delta$ and the GVD coefficient $\beta_D$. Indeed, a higher GVD coefficient and a lower detector time jitter---so that time-bin duration may be decreased---might increase HD-QKD's secret-key rate. The secret-key rates shown in Fig.~~\ref{allfigure} have already presumed a bin duration limited by state-of-the-art detector time jitter, but the $\beta_D$ value used is achievable with commercial devices~\cite{lee2014entanglement}.  Increasing the GVD coefficient without changing the other system parameters, however, does not always increase the secret-key rate. In particular, \eqref{securekey} shows that a $K$-fold increase in $\beta_D$ increases secret-key length by $\underline{n_{\sT,01}}\log_2(K)$, \emph{if} there is no offsetting increase in the error rate between Alice and Bob's raw keys, as quantified by the $\gamma(d_0 + \Delta)$ term in~\eqref{Eq:Hmax}.  Our numerical evaluation of the secret-key rate at zero distance versus $\beta_D$---using the other parameters from Table~\ref{Tab:exp:parameters} and the $d_0 = 2$ threshold code distance employed in Fig.~\ref{allfigure}---verifies this insight, see Fig.~\ref{dispersion}.  Here we see the secret-key rate initially increasing linearly with increasing $\log_{10}(\beta_D)$, until it saturates and begins to decrease.  Saturation occurs because our protocol requires $d_0 > d_{\rm min}$ for there to be a positive secret-key rate, and the minimum threshold code distance increases with increasing $\beta_D$, as shown in Appendix~\ref{thresholdmodeling}.  So, the secret-key rate saturation and decay in Fig.~\ref{dispersion} results from the $d_0$ increases that are required at high $\beta_D$ values.  That said, Fig.~\ref{dispersion}  still shows that the highest key rate, 17\,Mbit/s, is  realized with the experimentally feasible $\beta=2\times10^6\,$ps$^2 $~\cite{lee2014entanglement}, and we have found that the maximum distance for a non-zero secret-key rate is then 210\,km.

\section{Summary}\label{Conclusion}

We have reported the general-attack security analysis for the time-energy HD-QKD protocol in the finite-key regime by combining the entropic uncertainty-relation security analysis of CV-QKD with the decoy-state technique from DV-QKD. In particular, we derived a new entropic uncertainty relation for the time and conjugate-time operators using optical dispersion transformations. This result validates the difference between the uncertainty relation of time and conjugate-time operators and that of conventional maximally-incompatible operators, such as position and momentum. With the new uncertainty bound, we showed that under the most powerful attacks time-energy HD-QKD can produce a higher PIE than conventional decoy-state BB84 and CV-QKD, and still tolerate long-distance fiber transmission. We also showed that optimizing the HD-QKD protocol's GVD coefficient enables realizing a $17$\,Mbit/s secret-key rate at zero distance and a 210\,km maximum secure-transmission distance, the latter being comparable to that of state-of-the-art decoy-state BB84. We expect this finding will provide theoretical support for optimizing HD-QKD implementations. Our results constitute an important step toward the unified understanding of distinct QKD schemes that is needed for development of practical long-distance high-rate quantum communication.

\section*{Acknowledgment}
The authors thank Zheshen Zhang, Catherine Lee, Darius Bunandar, and Franco Wong for many helpful discussions. We acknowledge support from ONR grant number N00014-13-1-0774 and AFOSR grant number FA9550-14-1-0052.  F. Xu acknowledges support from an NSERC postdoctoral fellowship.

\begin{appendix}
\section{Protocol}\label{protocalAPP}

\begin{description}
  \item[a. Preliminaries]  Before contacting Bob, Alice makes measurements on her trusted spontaneous parametric down-conversion (SPDC) source of time-energy entangled biphotons to determine the coherence time of the pulsed pump field $\sigma_{\rm coh}$, the biphoton correlation time $\sigma_{\rm cor}$, and the SPDC intensities $\{\mu_1,\mu_2,\mu_3\}$, i.e., the mean photon-pairs generated per pump pulse with different pump powers.  Then, Alice and Bob use a pre-shared key to authenticate each other, after which they negotiate parameters to be employed during the protocol run.

  \item[b. Biphoton preparation and distribution] Alice pumps her SPDC source at a clock rate (repetition rate) $ R_{\mathrm{rep} }$. For each pump pulse, Alice prepares a time-energy entangled state within a $T_f$-duration ($T_f \approx \sigma_{\rm coh}$) frame centered on the peak of the pump pulse.  She sends one photon to Bob via a quantum channel (e.g., an optical fiber) and retains the companion photon for her own measurements. To implement decoy states~\cite{Wang2003,Lo2005,DariusPRA2015}, Alice randomly pumps the SPDC source to select intensities $\mu_k \in \{\mu_1,\mu_2,\mu_3 \}$ with probabilities $p_k \in \{p_{\mu_1},p_{\mu_2},p_{\mu_3}\}$.

  \item[c. Measurement phase] For each frame,  Alice and Bob select their measurement basis at random and independently from $\{\mathsf{T},\mathsf{W}\}$ with probabilities $\{q,1-q\}$ and perform measurements in their chosen bases. Their  $\mathsf{T}$-basis measurements are made using time-resolved single-photon detectors with a temporal resolution set primarily by the detectors' time jitter, $\sigma_{\rm jit}$~\cite{zhong2015photon}. They sort their data into time bins of duration $\delta$, where  $\sigma_{\rm cor} \ll \sigma_{\rm jit} < \delta \ll \sigma_{\rm coh}$, that will generate $\log_2(T_f/\delta)$ raw-key bits when they both obtain $\mathsf{T}$-basis photon detections in the same frame. Their $\mathsf{W}$-basis measurements are realized by means of dispersive optics and single-photon detection~\cite{mower2013high}, i.e., they pass their photons through normal and anomalous group-velocity dispersion (GVD) elements, respectively, measure them with time-resolved single-photon detectors, and then sort that data into duration-$\delta$ time bins.

  \item[d. Basis reconciliation] Alice and Bob announce their measurement bases over an authenticated public channel and discard all measurement results for frames in which they measured in different bases.  They are then left with detection-time coincidence measurements of $n_{\mathsf{T}}$ ($n_{\mathsf{W}}$) frames in which they both used the $\mathsf{T}$ ($\mathsf{W}$) basis and both obtained one photon detection.

  \item[e. Decoy-state processing] Alice announces her SPDC intensity choice for each frame. Alice and Bob thus identify sets  $\cT_{\mu_k}$ and $\cW_{\mu_k}$ for $\mu_k \in \{\mu_1,\mu_2,\mu_3 \}$, in which they have both made $\mathsf{T}$-basis or $\mathsf{W}$-basis measurements when Alice's SPDC source intensity was $\mu_k$.  They repeat their quantum communication, i.e., steps (b)--(e), until the cardinality of these sets satisfies: $|\cT_{\mu_k}| \geq n_{\sT,\mu_k}$ and $|\cW_{\mu_k}| \geq n_{\sW,\mu_k}$, where $\{n_{\sT,\mu_k},n_{\sW,\mu_k}\}$ are pre-chosen values that ensure sufficient quality in the ensuing parameter estimation steps. Note that $n_{\mathsf{T}}=\sum_{\mu_{k}} n_{\sT,\mu_k}$.  Next, they publicly announce their $\mathsf{W}$-basis detection times \{$t_{a,j,\mu_k}^{w}$, $t_{b,j,\mu_k}^{w}$\} for each SPDC intensity, where $a$, $b$ denote Alice and Bob, $j$ indexes the frame, and each detection-time value is relative to the peak of its associated pump pulse.  After that, they compute these detection times' mean-squared differences for each $\mu_k$, viz., $\sigma^2_{\rm cor,\sW, \mu_k}=\sum_j (t_{a,j,\mu_k}^{w}-t_{b,j,\mu_k}^{w})^2/n_{\sW,\mu_k}$.  By virtue of their use of normal and anomalous GVD elements,  $\sigma^2_{\rm cor,\sW, \mu_k}$ can be used to find the anti-correlation between the detunings from the SPDC outputs' center frequencies of the single-photon pairs (i.e., biphotons) that Alice and Bob detected in their $\mathsf{W}$-basis measurements when Alice's SPDC intensity was $\mu_k$~\cite{mower2013high}, see Appendix~\ref{App:decoy}.

  \item[f. Parameter estimation] Alice and Bob use only their $\mu_1$ data for secret-key generation, while they use their $\mu_2$ and $\mu_3$ data for parameter estimation. Alice and Bob use their $\mathsf{T}$-basis data to estimate $n_{\sT,0}$, the number of frames out of their $n_{\mathsf{T},\mu_1}$ that are due to vacuum coincidences (either Alice or Bob did not detect a photon),  and $n_{\sT,1}$, the number of frames out of their $n_{\mathsf{T},\mu_1}$ that are due to single-pair coincidences (Alice and Bob each detected one photon).  They use their $\mathsf{W}$-basis data to estimate $d_{\sW,1}$, the $\text{L}_{1}$ distance between their detected photons' frequency detunings (after accounting for their anti-correlation) that is due to single-pair coincidences~\cite{DariusPRA2015,lim2014concise} (see Appendix.~\ref{App:decoy}). Finally, they check that $d_{\sW,1}$ is less than $d_{0}$, where $d_{0}$ is a predetermined threshold (see Appendix.~\ref{thresholdmodeling}). If this condition is not met, they abort the protocol. Otherwise they proceed to the protocol's next step.

  \item[g. Key generation and error correction] Alice and Bob use their $\mathsf{T}$-basis data to generate raw keys $(\rv{X}_{\rm A},\rv{X}_{\rm B})$ from the frames in which Alice's SPDC intensity was $\mu_1$. Each frame used in generating these raw keys contains $\log_2(T_f/\delta)$ bits.  Alice and Bob perform error correction on their raw keys using an algorithm with reconciliation efficiency $\beta_e\leq 1$~\cite{zhou2013layered}. This procedure reveals at most $\leak$ bits of information to Eve. Next, to ensure that they have shared identical keys, Alice and Bob perform key verification using a two-universal hash function that publishes $\lceil \log_2(1/\eps_\tn{hash})\rceil$ bits of information, with $\eps_{\tn{hash}}$ being the probability that a pair of non-identical keys passes the test.

  \item[h. Calculation of secret-key length] Using the results from (f) and (g), Alice and Bob calculate the secret-key length $\ell$. If $\ell$ is negative, they abort the protocol. Otherwise, they apply another (different) two-universal hash function (for privacy amplification) to their error-corrected raw keys to produce the length-$\ell$ secret keys, $\rv{K}_{\rm A}$ and $\rv{K}_{\rm B} $.
\end{description}

\section{Time-frequency uncertainty relation for dilated measurements}\label{InfiniteOverlap}
To compare with the overlap developed in the main text, we derive the overlap with dilation in this appendix. Instead of the frequency domain, it is now more convenient to work in the time domain, using $\ket{t}=a^{\dagger}(t)\ket{0}$ to denote the single-photon localized at time $ t $ that satisfies the orthonormality condition $\braket{t_1}{t_2} = \delta(t_1 - t_2)$. The single-photon Hilbert space is simply $\cH = L^2(T) $, i.e., the space of square-integrable, complex-valued functions on the time-domain  $t \in T$.  We evaluate the overlap under dilation~\cite{werner1987,kiukas2012} when $ T=(-\infty, \infty)$. In this case, the POVMs $T(t)$ and $W(t)$ for the time and conjugate-time measurements are projection valued~\cite{mower2013high}:
\begin{align}\label{continuousT}
&T(t) =\ket{t}\bra{t},\\
&W(t) = UT(t) U^\dagger.\label{continuousD}
\end{align}
Here, $W(t)$ is obtained from $T(t)$ via the unitary transformation
\begin{eqnarray}
U= \frac{1}{\sqrt{\pi \beta_D}} \int_{-\infty}^\infty\!{\rm d}t_1\int_{-\infty}^{\infty}\!{\rm d}t_2\,e^{-i(t_1-t_2)^2/\beta_D} \ket{t_1}\bra{t_2}.
\end{eqnarray}
The associated time and conjugate-time observables are then
\begin{eqnarray}
O_t&=&\int_{-\infty}^{\infty}\!{\rm d}t\,  t\ket{t}\bra{t},\\[.05in]
D_t&=&\frac{1}{\pi \beta_D}\int_{-\infty}^\infty\!{\rm d}t\,t\!\int_{-\infty}^\infty\!{\rm d}t_1\!\int_{-\infty}^{\infty}\!{\rm d}t_2\, e^{-i(t_1^2-t_2^2)/\beta_D}  \nonumber \\[.05in]
&\times& e^{2i(t_1-t_2)t/\beta_D}\ket{t_1}\bra{t_2}.
\end{eqnarray}
The conjugate-time observable can be further simplified as follows:
\begin{widetext}
\begin{align}
D_t&= \frac{1}{\pi \beta_D}\int_{-\infty}^{\infty}\!{\rm d}t_1\int_{-\infty}^\infty\!{\rm d}t_2\, e^{-i(t_1^2-t_2^2)/\beta_D} \!\left(\frac{\partial}{\partial t_1}\int_{-\infty}^{\infty}\!{\rm d}t\,\frac{\beta_D}{2i} e^{2i(t_1-t_2)t/\beta_D}\right)\!\ket{t_1}\bra{t_2},\\[.05in]
&= \frac{\beta_D}{2i}\int_{-\infty}^\infty\!{\rm d}t_1\,e^{-it_1^2/\beta_D}|t_1\rangle\!\left(\frac{\partial}{\partial t_1}\int_{-\infty}^\infty\!{\rm d}t_2\,e^{it_2^2/\beta_D}\delta(t_1-t_2)\langle t_2|\right) \\[.05in]
  &= \int_{\infty}^\infty\!{\rm d}t_1\,t_1|t_1\rangle\langle t_1| + \frac{\beta_D}{2i}\int_{-\infty}^{\infty}\!{\rm d}t_1\, |t_1\rangle \,\frac{\partial}{\partial t_1}\langle t_1| \\[.05in]
  &= O_t + \pi\beta_DO_\omega, \label{cvRelation}
\end{align}
\end{widetext}
where $O_\omega = \int_{-\infty}^\infty\!\ddo{\omega}\,\omega |\omega\rangle\langle \omega|$ is the conventional unbounded-frequency observable that is maximally incompatible with the time observable. It immediately follows that
\begin{equation}
[O_t, D_t] =i\pi\beta_D.
\end{equation}

Finally, using the overlap result for maximally-incompatible observables~\cite{Fabian2015}, we obtain
\begin{eqnarray}
c(\delta, \beta_D)= \frac{\delta^2}{2\pi^2 \beta_D},
\end{eqnarray}
for the dilated measurements.  Compared with the overlap derived with non-projective POVM in Sec.~\ref{secUncertainty}, this overlap is slightly smaller, and thus offers a weaker bound on the uncertainty relation. In the paper we therefore used the non-dilated overlap in bounding the secret-key length.

\section{Decoy states with finite keys}~\label{App:decoy}
A decoy-state method for HD-QKD in the asymptotic regime was previously derived in~\cite{DariusPRA2015}. Here, based on~\cite{lim2014concise}, we extend the work in~\cite{DariusPRA2015} to the finite-key case against \emph{general} attacks (i.e., without any assumptions on the statistical distributions). We presume that Alice randomly chooses between three intensity levels, $\mu_1, \mu_2$ and $\mu_3$ for her SPDC source. Let $s_{\sT,n}$~be the number of frames in which Alice and Bob both measure in the $\sT$ basis and Alice's source has emitted $n$ biphotons in each frame, so that $n_{\sT} = \sum_{n=0}^\infty s_{\sT,n}$ is the total number of frames in which Alice and Bob both made $\sT$-basis measurements. In the asymptotic regime, $n_{\sT,\mu_k}$, the number frames in which Alice's source intensity was $\mu_k$ and she and Bob made $\sT$-basis measurements, approaches its ensemble-average value, namely
\[
n_{\sT,\mu_k} \rightarrow n^*_{\sT,\mu_k}= \sum_{n=0}^\infty p_{\mu|\sn}(\mu_k|n)s_{\sT,n},~\mbox{for $\mu_k   \in \{\mu_\rs,\mu_\rd,\mu_\rdd\}$},
\]
where $p_{\mu|\sn}(\mu_k|n)$ is the conditional probability of Alice's source emitting $\sn = n$ biphotons in a frame, given its source intensity was $\mu=\mu_k$. For finite sample sizes, Hoeffding's inequality for independent events~\cite{hoeffding1963probability} implies that $n_{\sT,\mu_k}$ will satisfy
\be \label{A:eqn:1}
\left|n^*_{\sT,\mu_k}-n_{\sT,\mu_k} \right| \leq \zeta(n_{\sT},\eps_1),
\ee
with probability at least $1-2\eps_1$, where $\zeta(n_{\sT},\eps_1):=\sqrt{n_{\sT}\log(1/\eps_1)/2}$. Note that the deviation term $\zeta(n_\sT,\eps_1)$ is the same for all $\mu_k$. Inequality~\eqref{A:eqn:1} allows us to establish a relation between the asymptotic values $\{n^*_{\sT,\mu_k}\}$ and the observed values $\{n_{\sT,\mu_k}\}$. More precisely, we have the following bounds for finite-key analysis:
\begin{eqnarray} \label{A:eqn:7}
 n_{\sT,\mu_k}^* &\leq& n_{\sT,\mu_k}+\zeta(n_\sT ,\eps_1)=:\overline{n_{\sT,\mu_k}}, \\ \label{A:eqn:8}
 n_{\sT,\mu_k}^* &\geq& n_{\sT,\mu_k}-\zeta(n_\sT ,\eps_1)=:\underline{n_{\sT,\mu_k}}.
\end{eqnarray}

\subsubsection{Lower-bound on the number of vacuum coincidences, $\underline{n_{\sT,0}}$}
The following lower bound on $s_{\sT,0}$ was derived in Ref.~\cite{lim2014concise}: 
\be \label{A:eqn:4}
s_{\sT,0} \geq \underline{s_{\sT,0}} = \frac{\tau_0}{(\mu_\rd-\mu_\rdd)}\left(\frac{\mu_\rd e^{\mu_\rdd} \underline{n_{\sT,\mu_\rdd}}}{p_{\mu_\rdd}}-\frac{\mu_\rdd e^{\mu_\rd}  \overline{n_{\sT,\mu_\rd}}}{p_{\mu_\rd}}\right).
\ee
Using this result we obtain the lower bound on the number of vacuum coincidences when Alice's source intensity is $\mu_1$ given by
$$
n_{\sT,0}\geq \underline{n_{\sT,0}}=\underline{s_{\sT,0}}p_{\mu|\sn}(\mu_1|0).
$$

\subsubsection{Lower bound on the number of single-pair coincidences, $\underline{n_{\sT,1}}$}
The following lower bound on $s_{\sT,1}$ was derived in 
Ref.~\cite{lim2014concise}:
\begin{multline} \label{A:eqn:5}
s_{\sT,1} \geq \underline{s_{\sT,1}} = \frac{\mu_\rs \tau_1 }{\mu_\rs(\mu_\rd-\mu_\rdd)-(\mu_\rd^2-\mu_\rdd^2)}
\Bigg[  \frac{ e^{\mu_\rd}  \underline{n_{\sT,\mu_\rd}}}{p_{\mu_\rd}} \\ -\frac{ e^{\mu_\rdd} \overline{n_{\sT,\mu_\rdd}}}{p_{\mu_\rdd}} + \frac{\mu_\rd^2-\mu_\rdd^2}{\mu_\rs^2} \left(    \frac{s_{\sT,0}}{\tau_0}-\frac{e^{\mu_\rs}\overline{n_{\sT,\mu_\rs}}}{p_{\mu_\rs}} \right)                \Bigg].
\end{multline}
Using this result we obtain the lower bound on the number of single-pair coincidences when Alice's source intensity is $\mu_1$ given by
$$
n_{\sT,1}\geq \underline{n_{\sT,1}}=\underline{s_{\sT,1}}p_{\mu_1|1}-\zeta(\underline{s_{\sT,1}}p_{\mu|\sn}(\mu_1|1),\eps_2),
$$
with probability at least $1-2\epsilon_2$.

\subsubsection{Upper bound on the $\text{L}_{1}$ distance of single-pair coincidences, $\overline{d_{\sW,1}}$}
After the non-local dispersion cancellation that occurs when Alice and Bob both make $\mathsf{W}$-basis measurements on the same frame, their mean-square time difference, $\sigma^2_{{\rm cor},\sW,\mu_k}$ for $\mu_k \in\{\mu_\rs,\mu_\rd,\mu_\rdd\}$, can be written as
\begin{align*}
\sigma^2_{{\rm cor},\sW,\mu_k}&=\frac{p_{\mu|\sn}(\mu_k|1)s_{\sW,1}}{n_{\sW,\mu_k}}\sigma^2_{{\rm cor},\sW,1}\\
&+\left(1-\frac{p_{\mu|\sn}(\mu_k|1)s_{\sW,1}}{n_{\sW,\mu_k}}\right)\sigma^2_{{\rm cor},\sW,m},
\end{align*}
where $\sigma^2_{{\rm cor},\sW,1}$ and $\sigma^2_{{\rm cor},\sW,m}$ are the mean-squared differences due to single-pair and multiple-pair coincidences including \emph{all} source intensities, and $s_{\sW,1}$ is the number of frames in which Alice and Bob both measure in the $\sW$ basis given that Alice's source has emitted $1$ biphoton in each frame. Then, we have
\begin{eqnarray}
\lefteqn{n_{\sW,\mu_\rd}\sigma^2_{{\rm cor},\sW,\mu_\rd} - n_{\sW,\mu_\rdd}\sigma^2_{{\rm cor},\sW,\mu_\rdd}= }\nonumber\\[.05in]
&& s_{\sW,1}\sigma^2_{{\rm cor},\sW,1}[p_{\mu|\sn}(\mu_\rd|1)-p_{\mu|\sn}(\mu_\rdd|1)]  \nonumber \\[.05in]
&&+\,\,\sigma^2_{{\rm cor},\sW,m}[n_{\sW,\mu_\rd}-n_{\sW,\mu_\rdd} \nonumber\\[.05in]
&&+\,\,p_{\mu|\sn}(\mu_3|1)s_{\sW,1}-p_{\mu|\sn}(\mu_\rd|1)s_{\sW,1}],
\end{eqnarray}
where the $\sigma^2_{{\rm cor},\sW,m}$ term on the right is non-negative for $\mu_\rd > \mu_\rdd$. Dropping the $\sigma^2_{{\rm cor},\sW,m}$ term, the preceding result can be rearranged to provide the lower bound
\be \label{A:eqn:10}
\sigma^2_{{\rm cor},\sW,1} \leq \overline{\sigma^2_{{\rm cor},\sW,1}}= \frac{\overline{n_{\sW,\mu_\rd}}\sigma^2_{{\rm cor},\sW,\mu_\rd}-\underline{n_{\sW,\mu_\rdd}}\sigma^2_{{\rm cor},\sW,\mu_\rdd}}{\underline{s_{\sW,1}}(p_{\mu|\sn}(\mu_\rd|1)-p_{\mu|\sn}(\mu_\rdd|1))},
\ee
where the $s_{\sW,1}$ lower bound, $\underline{s_{\sW,1}}$, can be derived using the same method employed in~\cite{lim2014concise} to obtain inequality~(\ref{A:eqn:5}). Our upper bound on the $\text{L}_{1}$ distance of single-pair coincidences is then
\be \label{A:eqn:11}
\overline{d_{\sW,1}}=\sqrt{ \frac{2}{\pi}\overline{\sigma^2_{{\rm cor},\sW,1}}},
\ee
where the $\sqrt{2/\pi}$ factor arises from relating $\text{L}_1$ distance to the mean-squared difference of jointly Gaussian random variables.

\section{Theoretical model for the threshold, $d_{0}$\label{thresholdmodeling}}
To find the threshold, $d_0$, for the mean-squared difference between Alice and Bob's single-pair $\sW$-basis measurements beyond which Alice and Bob will abort the QKD protocol, we start from the time and frequency wave-functions for the biphoton emission when Alice's SPDC source is pumped by a pulse centered at time $t=0$ \cite{zhang2014unconditional}, i.e.,
\begin{align}
\psi(t_S,t_I) &= \frac{\exp(-t_-^2/4\sigma^2_{\rm coh} -t_+^2/4\sigma^2_{\rm cor} -i\omega_Pt_+)}{\sqrt{2\pi\sigma_{\rm coh}\sigma_{\rm cor}}},
\\[.05in]
\Psi(\omega_S,\omega_I) &= \frac{\exp(-\omega_-^2\sigma^2_{\rm cor}/4 -4\omega_+^2\sigma^2_{\rm coh})}{\sqrt{\pi/2\sigma_{\rm coh}\sigma_{\rm cor}}}.
\end{align}
Here: $t_S$ and $t_I$ denote the times of the biphoton's signal and idler photons, and $\omega_S$ and $\omega_I$ denote their frequencies;  $t_+ := (t_S+t_I)/2$, $t_- := t_S-t_I$, $\omega_+ := (\omega_S + \omega_I)/2$, and $\omega_- := \omega_S -\omega_I$; and we have assumed that Alice's source is phase matched at frequency degeneracy for its pump's $\omega_P$ center frequency.

When both Alice and Bob choose the conjugate-time basis, they send their photons into normal and anomalous group-velocity dispersion elements whose dispersion coefficients have common magnitude $\beta_D$ but opposite signs.  After the propagation through the dispersive elements at Alice and Bob's terminal, the frequency wave-function becomes
\begin{align}
\lefteqn{\Psi_D(\omega_S,\omega_I) = } \nonumber \\[.05in]
&& \hspace*{-.15in}\frac{\exp[-\omega_-^2\sigma^2_{\rm cor}/4 -4\omega_+^2\sigma^2_{\rm coh} +i\beta_D/4 (\omega_S^2- \omega_I^2) ]}{\sqrt{\pi/2\sigma_{\rm coh}\sigma_{\rm cor}}},
\end{align}
from which the associated time wave-function can be found via
\begin{eqnarray}
\lefteqn{\psi_D(t_S,t_I) = } \nonumber \\[.05in]
&& \hspace*{-.2in}\frac{1}{2\pi}\int_{\infty}^\infty\!{\rm d}\omega_S\int_{-\infty}^\infty\!{\rm d}\omega_I\,\Psi_D(\omega_S,\omega_I)
e^{-i(\omega_St_S + \omega_It_I)}.
\end{eqnarray}
The $\sW$-basis mean-squared time difference in the absence of Eve is therefore
\begin{align}
\sigma^2_{{\rm cor},\sW} &= \int_{-\infty}^\infty\!{\rm d}t_S\int_{-\infty}^\infty\!{\rm d}t_I\,(t_S-t_I)^2|\psi_D(t_S,t_I)|^2 \\[.05in]
&= \frac{\sigma^2_{\rm coh}\sigma^2_{\rm cor} + (\beta_D /4)^2}{\sigma^2_{\rm coh}},\\
&= \sigma^2_{\rm cor} +\frac{\beta_D^2}{16\sigma^2_{\rm coh}}
\end{align}

This correlation time measures how strongly Alice and Bob's single photons are correlated in the conjugate-time basis in the absence of Eve. Subsequently, we find the minimum $ L_1 $ distance for conjugate-time measurement outcomes without any third-party interference to be
\begin{align}
d_{\rm min}= \sqrt{\frac{16\sigma^2_{\rm coh}\sigma_{\rm cor}^2 +\beta_D^2}{8\pi\sigma_{\rm coh}^2\delta^2}},
\end{align}
where the $1/\delta$ factor normalizes the root-mean-square time difference into time bins and the $\sqrt{2/\pi}$ factor converts root-mean-square bin difference into $L_1$ distance.  The $d_0$ that determines when Alice and Bob will abort their QKD protocol thus should be bigger than $ d_{\rm min} $ in order to have non-zero key rate. In our performance evaluation, whose results are shown in Fig.~\ref{allfigure}, we chose $ d_0= 2$.  This value is well above this $d_{\rm min}$ lower bound for the parameter values given in Table~\ref{Tab:exp:parameters}.

\end{appendix}


\begin{thebibliography}{99}
\bibitem{Footnote1}The clock rate is assumed to be the inverse of three times the pump coherence time.
\end{thebibliography}
\end{document}